\documentclass[10pt,conference,letterpaper]{IEEEtran}
\IEEEoverridecommandlockouts
\usepackage{cite}
\usepackage{amsmath,amssymb,amsfonts}
\usepackage{graphicx}
\usepackage{algorithmicx}
\usepackage{booktabs}
\usepackage[noend]{algpseudocode}
\usepackage{subfloat}
\usepackage{textcomp}
\usepackage{xcolor}
\usepackage{enumitem}
\usepackage{array}
\usepackage[font=footnotesize]{caption}
\usepackage{footnote}
\usepackage{url}
\usepackage{balance}
\usepackage{multirow}
\usepackage{hyperref}
\usepackage{gensymb}
\usepackage[caption=false,font=footnotesize]{subfig}
\def\BibTeX{{\rm B\kern-.05em{\sc i\kern-.025em b}\kern-.08em
    T\kern-.1667em\lower.7ex\hbox{E}\kern-.125emX}}
\newcolumntype{b}{X}
\newcolumntype{s}{>{\hsize=.5\hsize}X}

\renewcommand{\topmargin}{-.7in}

\makeatletter
\newcommand\fs@spaceruled{\def\@fs@cfont{\bfseries}\let\@fs@capt\floatc@ruled
  \def\@fs@pre{\vspace*{0.2cm}\hrule height.8pt depth0pt \kern2pt}%
  \def\@fs@post{\kern2pt\hrule\relax}%
  \def\@fs@mid{\kern2pt\hrule\kern2pt}%
  \let\@fs@iftopcapt\iftrue}
\makeatother

\begin{document}
%\vspace*{0.05cm}
%\title{SAT-TRAIN: A Large Trace-Based Dataset of Real-World Starlink Measurements}
\title{WetLinks: a Large-Scale Longitudinal Starlink Dataset with Contiguous Weather Data}

\author{ 
Dominic Laniewski\textsuperscript{1}, Eric Lanfer\textsuperscript{1}, Bernd Meijerink\textsuperscript{2}, Roland van Rijswijk-Deij\textsuperscript{2}, Nils Aschenbruck\textsuperscript{1}\\

\textsuperscript{1}Osnabrück University - Institute of Computer Science, Osnabrück, Germany\\
\textsuperscript{2}University of Twente - Design and Analysis of Communication Systems Group, Enschede, The Netherlands\\
\{laniewski, lanfer, aschenbruck\}@uos.de, \{bernd.meijerink, r.m.vanrijswijk\}@utwente.nl\\
}

%\author{Anonymous authors}

\maketitle

\begin{abstract}
Low Orbit Satellite (LEO) networks such as Starlink promise Internet access everywhere around the world.
In this paper, we present \emph{WetLinks} - a large and publicly available trace-based dataset of Starlink measurements. The measurements were concurrently collected from two European vantage points over a span of six months. Consisting of approximately 140,000 measurements, the dataset comprises all relevant network parameters such as the upload and download throughputs, the RTT, packet loss, and traceroutes. We further augment the dataset with concurrent data from professional weather stations placed next to both Starlink terminals. 
Based on our dataset, we analyse Starlink performance, including its susceptibility to weather conditions.
%Based on our dataset, we analyse Starlink performance as well as the impact of weather conditions on throughput. 
We use this to validate our dataset by replicating the results of earlier smaller-scale studies. We release our datasets and all accompanying tooling as open data. To the best of our knowledge, ours is the largest Starlink dataset to date.
\begingroup%
\renewcommand\thefootnote{}
\footnote{Dominic Laniewski and Eric Lanfer are co-first authors}%
\addtocounter{footnote}{-1}
\endgroup%
\end{abstract}


\IEEEpeerreviewmaketitle

% !TEX root = main_ieee.tex

\section{Introduction}

In networking research, new ideas and approaches are commonly developed and tested via simulations. These simulations often require data on realistic network conditions such as throughputs, packet loss, and latency.
%In those, often realistic network conditions such as throughputs, packet loss, and latencies need to be implemented.
This data can either be generated using network simulators such as \textit{ns3}\footnote{\url{https://www.nsnam.org}}, by using models such as the Gilbert-Elliot model \cite{hasslinger2008gilbert} for bursty packet loss, or by replaying real-world traces.
The benefit of using real-world traces is that they represent realistic network conditions.
Trace-based evaluations allow other researchers not only to replicate presented results, but also to benchmark their own approaches against existing solutions under the same network conditions, leading to an overall increased quality of research.
The need for real-world network traces is amplified by the rise of machine learning (ML) methods in many networking research areas, such as rate adaptation for video-streaming~\cite{kan2021rapt360} and point cloud streaming~\cite{laniewski2021potential}, traffic prediction \cite{fauzi2022mobile}, and intrusion detection~\cite{shone2018deep}.
ML-based solutions, especially ones that apply deep learning (DL) techniques, require large training datasets to achieve a high level of performance and to avoid overfitting of the models to certain network conditions. %While such datasets usually need to be domain-specific, many applications, for example, rate adaptation \cite{laniewski2021potential}, need large datasets of real-world network traces to avoid an overfitting of the models to certain network conditions.

%In general, there is a lack of publicly available large-scale longitudinal datasets of traces from real-world network measurements.
%In the past, datasets of 3G~\cite{riiser2013commute}, 4G~\cite{raca2018beyond}, and 5G~\cite{raca2020beyond} measurements have been published. However, since they have not been created with ML applications in mind, their size is limited.
Lately, the first studies of SpaceX's Starlink Low Earth Orbit (LEO) satellite were published~\cite{kassem2022browser,ma2023network,michel2022first}. Most studies published since Starlink opened for public access in 2020 focus on first impressions of achievable performance. While these studies made their datasets public, all of these datasets suffer from limitations. They are all: 1) limited in size, 2) have an {\it ad hoc} methodology, 3) lack a clear description of measurement conditions, such as equipment used, user load, etc., and 4) focus only on specific network parameters such as latency.
%These datasets often suffer from a combination of the following limitations:
%To the best of our knowledge, all these datasets suffer from a combination of the following limitations:
%\begin{itemize}
%    \item limited in size;
%    \item the measurements have not been conducted fully systematically;
%    \item they are not sufficiently described such as in terms of the conditions under which they were captured, the used equipment, and the user load;
%    \item they focus only on specific network parameters such as latency.
%\end{itemize}
In general, creating a large-scale longitudinal trace-based dataset of continuous Starlink measurements of all relevant network parameters is a costly project, as it requires exclusive use of a Starlink dish for the duration.

In this paper, we intend to remedy this by presenting \emph{WetLinks}: a large-scale longitudinal Starlink dataset. \emph{WetLinks} consists of six months (Oct.~'23 -- Mar.~'24) of orchestrated Starlink measurements at 3-minute intervals from two European sites -- in Osnabrück%\footnote{Redacted for double-blind review}
~(GER) and Enschede (NL). With a distance as the crow flies of $\pm$80~km between the measurement stations, both dishes are connected to the 53\degree~orbit and likely connect to the same satellites in sequence. The measurements include all relevant network parameters such as the download and upload throughputs, the Round-Trip-Time (RTT), the packet loss rate (PLR), traceroutes, as well as accurate weather data from reference weather stations placed in direct physical proximity to the dishes.
Our contributions are as follows:
%\begin{itemize}[noitemsep,nosep,topsep=0pt,labelindent=0em,leftmargin=*]
\begin{itemize}[itemsep=1pt,topsep=1pt,labelindent=0.25em,leftmargin=*]
    \item We make \emph{WetLinks} publicly available. It consists of approximately 80,000~measurements from Osnabrück and 60,000~from Enschede. To the best of our knowledge, this is the largest and most complete Starlink dataset to date. %. At the time of publication, it will consist of data from six months of measurements, totaling to approx. 80,000 measurements from Osnabrück, and 60,000 measurements from Enschede. To the best of our knowledge, our dataset is currently the largest publicly available dataset of Starlink measurements, containing high-frequency measurements for every relevant network parameter.
    %\item We pair our measurements with accurate weather data directly captured next to the dishes.
    %\item We validate our dataset by showing that our analysis of Starlink's performance, including analysis of the impact of weather conditions, is consistent with earlier studies.
    \item We pair our measurements with accurate weather data directly captured next to the dishes and analyze the impact of weather conditions on Starlink's performance.
    \item We validate our dataset by replicating the results of earlier smaller-scale studies.
    \item We release companion tooling that merges different data sources (performance measurements, weather data, \ldots) together with the dataset.
    %\item We make our analysis scripts publicly available to allow a full replicability of our results. 
\end{itemize}

The remainder of this paper is structured as follows.
Sec.~\ref{sec:background} gives background on Starlink and its radio communications. In Sec.~\ref{sec:related_work} we discuss related work on existing studies of Starlink performance and weather impact on satellite communications.
Next, Sec.~\ref{sec:setup} describes our measurement setup. %Then, we present our dataset in Sec.~\ref{sec:dataset}.
In Sec.~\ref{sec:analysis}, we validate our dataset by analysing performance and weather impact and comparing this against earlier studies. Finally, we conclude the paper in Sec.~\ref{sec:conclusion}.

% !TEX root = main_ieee.tex

\section{Background}\label{sec:background}

\subsection{Starlink Overview}
Starlink is a broadband Internet service provided by SpaceX. It relies on a Low Earth Orbit (LEO) satellite constellation initially (from 2018) orbiting at an altitude of 1,100\,km~\cite{fcc1838}, and since 2021 the orbits have been lowered to between 540\,km and 570\,km~\cite{fcc2148}.
%The company SpaceX provides broadband network access via a satellite network, called Starlink, operating on low earth orbits (LEO). These LEO satellites are operating at an altitude of $540$ km to $570$ km, before an FCC approval in 2021, satellites were at an altitude higher than $1,100$ km\cite{fcc1838}\footnote{FCC 21-48, April 2021}. 
So far, SpaceX launched 5,422~satellites, of which some 4,700~are operational~\cite{starlinkmega}. These are deployed in three orbits of varying density at inclinations of 53\degree, 70\degree, and 97.7\degree. The 53\degree\ orbit has the most satellites and provides most of the service. The 70\degree\ and 97.7\degree\ orbits have fewer satellites and mainly serve the Earth's polar regions~\cite{mohan2023multifaceted}. %The aim of this mega constellation of satellites is to provide global access to the Starlink network. 

Customers connect to the network via their user terminal, which is colloquially known as a ``Dishy''. Bidirectional communication between the terminal and the satellite makes use of $K_u$-Band beams. Communication between satellites and ground stations makes use of the $K_a$-Band.
%For the downlink from satellite to user terminal, $K_u$-Band beams are used. The terminal's uplink to the satellites is also realized using the $K_u$-Band. The Gateways, also called ground stations, operate in the $K_a$-Band.
Starlink has a reconfiguration interval of 15\,s, where the terminal can be rescheduled to another satellite and frequencies and routing resources can be reallocated~\cite{mohan2023multifaceted}.
SpaceX claims that downlink speeds up to 220\,Mbit/s and uplinks speeds up to 25\,Mbit/s are possible at latencies ranging from 25 to 60\,ms\cite{starlink-specs}. SpaceX offers \emph{Standard} and \emph{Priority} service plans. Priority traffic is given network precedence over standard traffic~\cite{starlink-fairuse} at times of high network load and congestion.

The Starlink network uses a one-hop bent-pipe routing approach if both the user terminal and the ground station are within line of sight of a single satellite~\cite{ma2023network}. %In this case, the satellite with the connected user terminal forwards the traffic directly down to the closest earth ground station without any further routing between the satellites \cite{ma2023network}.
In cases where one satellite does not cover both endpoints, inter-satellite links (ISL) using laser-beams can be utilized to form an extended multi-hop bent-pipe~\cite{mohan2023multifaceted}. Only newer generation satellites starting with version~1.5 support ISL.

\subsection{$K_{u}$- and $K_{a}$-band fundamentals}
As discussed above, Starlink uses the $K_{u}$- and $K_{a}$-band (as defined in~\cite{ieee521-2002}) for satellite to earth communications. The $K_{u}$-band is used for terminal to satellite communications, where frequencies between 10.7-12.7\,GHz are used for satellite-to-terminal and frequencies between 14.0-14.5\,GHz are used for terminal-to-satellite. % and has a frequency range of 12-18\,GHz, with a wavelength range of 2.5-1.67\,cm.% As mentioned in the section before, the end user terminal to satellite communication is realized using the $K_{u}$-band. For download, a frequency range of $10.7$ to $12.7$ GHz is used, and for uploads, the used frequency range is $14.0$ to $14.5$ GHz.
The $K_{a}$-band is then used for satellite to ground station communication with frequencies between 17.8-18.6\,GHz being used for satellite-to-ground, and frequencies between 27.5-29.1\,GHz and between 29.5-30\,GHz being used for ground-to-satellite.
%
%For the $K_{a}$-band, a wavelength range of $11.1-7.5$ mm at a Frequency range of $27-40$ GHz is standardized \cite{ieee521-2002}. Starlink implements the satellite to ground station (Gateways) communication in the $K_{a}$-band. As mentioned before, Starlink uses a wider ranger close to the standardized bands. The downlink, satellite to gateway, operates at $17.8$ to $18.6$ GHz and $18.8$ to $19.3$ GHz. The uplink is on $27.5$ to $29.1$ GHz and $29.5$ to $30$ GHz~\cite{fcc1838}. 

Since both bands run at frequencies higher than 10\,GHz, they are susceptible to tropospheric phenomena that affect signal propagation \cite{ippolito1989propagation,panagopoulos2004satellite}. The effects include, among others, attenuation due to precipitation and clouds. In this work, we combine weather data with throughput measurements on the terminal-to-satellite link to analyse the impact of precipitation and cloud attenuation on Starlink throughput.
%\begin{itemize}
%    \item Attenuation Due to Precipitation
%    \item Gaseous Absorption
%    \item Cloud Attenuation
%    \item Melting Layer Attenuation
%    \item Sky Noise Increase
%    \item Signal Depolarization
%    \item Tropospheric Scintillation
%    \item Intersystem Interference
%\end{itemize}

% noch auf Rain Attenuation weiter eingehen?
% We will not go into further detail on the single effects and refer to the previously mentioned literature. However, to quantify the weather influences on these two bands, we will discuss studies and related work, which are not necessarily Starlink related, on this topic in the next section.

 %Participants connect to the network via their user terminal. For the downlink from satellite to user terminal, $K_u$-Band beams are used in a frequency range of $10.7$ to $12.7$ GHz. The terminal’s uplink to the satellites is also realized using the $K_u$-Band, with a frequency range of $14.0$ to $14.5$ GHz. The Gateways, also called ground stations, operate in the $K_a$-Band. 
%Starlink has a reconfiguration interval of 15~s, where the terminal can be rescheduled to another satellite and frequencies and routing resources can be reallocated \cite{mohan2023multifaceted}.
%SpaceX itself notes that download speeds up to $220$ Mbps and uploads up to $25$ Mbps are possible at latencies ranging from $25$ to $60$ ms\footnote{https://www.starlink.com/legal/documents/DOC-1400-28829-70}. 

% !TEX root = main_ieee.tex

\section{Related Work} \label{sec:related_work}
%The related work can be split into two broad categories: Starlink-related measurements and analysis, and more general analysis of the weather impact on $K_{a}$ and $K_{u}$ band communication.

%\subsection{Starlink}
Since Starlink opened to the public in 2020, several studies have analyzed the performance and internal functionings through simulations and real-world measurements.

On the simulation side, \texttt{StarPerf}~\cite{lai2020starperf} and \texttt{Hypatia}~\cite{kassing2020exploring} simulate the network behavior of satellite constellations. %, allowing easy and cheap access to data.
\texttt{LeoEM}~\cite{cao2023satcp} can also capture the dynamics of LEO satellite networks, and can be used to analyze TCP behavior.

Most measurement studies look at performance aspects. Michel et al.~\cite{michel2022first} characterise Starlink performance in terms of throughput over TCP and QUIC, latency and packet loss. Their longest measurement campaign spans 5~months, the study does not consider weather influences. Kassem et al.~\cite{kassem2022browser} recruited volunteers that installed a browser extension and performed a measurement campaign over a period of 6~months. They augment this with a limited but more intensive measurement campaign from three locations using Raspberry Pi hosts connected to volunteer Starlink dishes for an unspecified amount of time. They study weather influence for one location using openly available data from a nearby weather station. Ma et al.~\cite{ma2023network} collect routing information and conduct performance measurements including for different types of applications, such as video streaming. They deploy 4~Starlink dishes and measure over a period of 7~months at irregular intervals. 
Izhikevich et al.\cite{izhikevich2023democratizing} conduct world-wide latency measurements over one month by actively probing publicly exposed devices at unspecified intervals that are connected via Starlink. 
Zhao et al.~\cite{zhao2023realtime} focus on real-time multimedia services, such as video-on-demand, live streaming and video conferencing. They note that performance is typically adequate for these services but may be impacted by satellite handovers and adverse weather in the form of thunderstorms. Pan et al.~\cite{pan2023measuring} also measure performance and perform traceroutes, focusing on analyzing Starlink's point-of-presence (POP) structure. Raman et al.~\cite{raman2023dissecting} compare Starlink against medium- and geostationary orbit (MEO/GEO) satellite networks finding that Starlink outperforms these older technologies. Garcia et al.~\cite{garcia2023multi} study Starlink performance with a specific focus on system-specific timing structures such as frequency scheduling and beam switching. Finally, Mohan et al.~\cite{mohan2023multifaceted} analyze M-Lab speed test data of Starlink measurements and compare this against RIPE Atlas measurements from probes with Starlink connectivity and measurements using their own Starlink dishes.

In contrast to previous studies discussed above, we are the first to provide a large and fine-grained dataset of continuous Starlink measurements at minute-intervals conducted over the course of six months from two European cities including high-precision weather data measured with reference weather stations placed directly next to the Starlink dishes. This allows for accurate analysis of the impact of weather conditions on Starlink performance. Furthermore, the open nature of our dataset paves the way for future studies that require large, continuous input datasets, such as studies that rely on machine learning approaches.

\section{Measurement Setup}\label{sec:setup}

\begin{figure}[t]
    \centering
    \includegraphics[width=0.8\columnwidth]{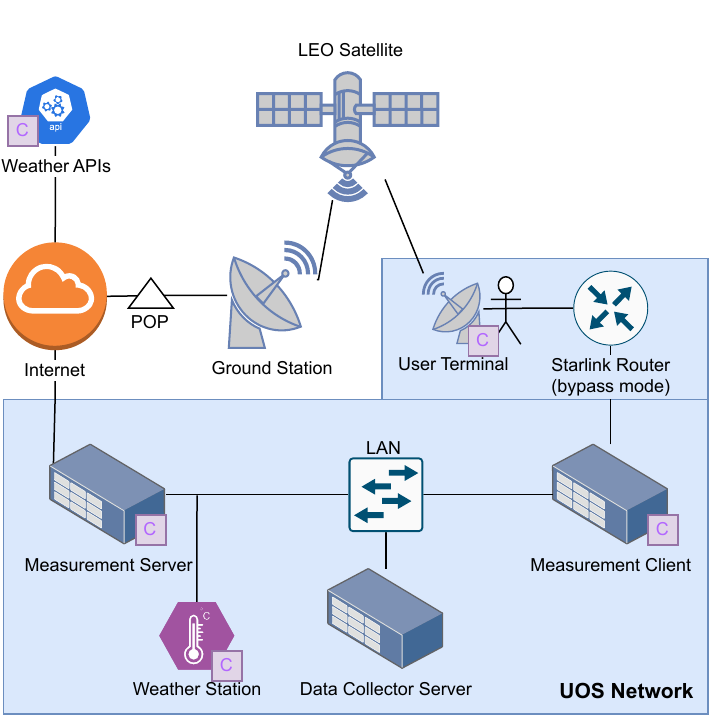}
    \caption{The Measurement Setup. The purple $C$ indicates data endpoints, that are collected by the data collection server.}
    \label{fig:general_setup}
\end{figure}

\begin{figure}[t]
    \centering
    \subfloat[Osnabrück]{
        \includegraphics[width=0.4\linewidth]{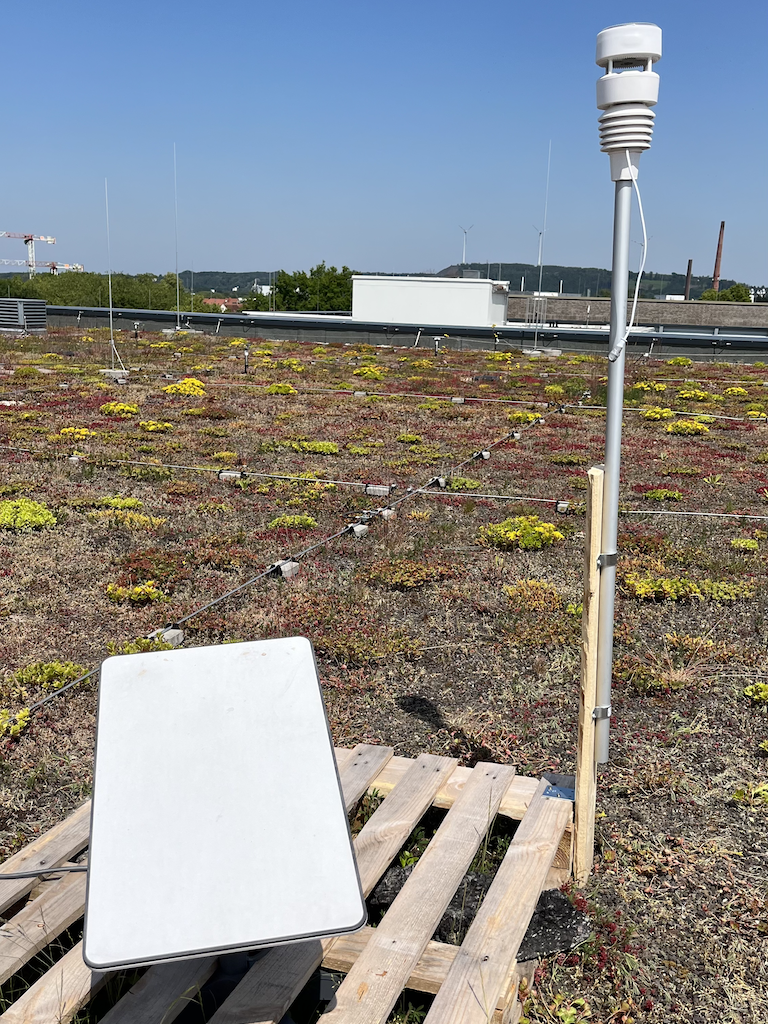}
        \label{fig:setup_Osnabrück}
    }
    \subfloat[Enschede]{
        \includegraphics[width=0.4\linewidth]{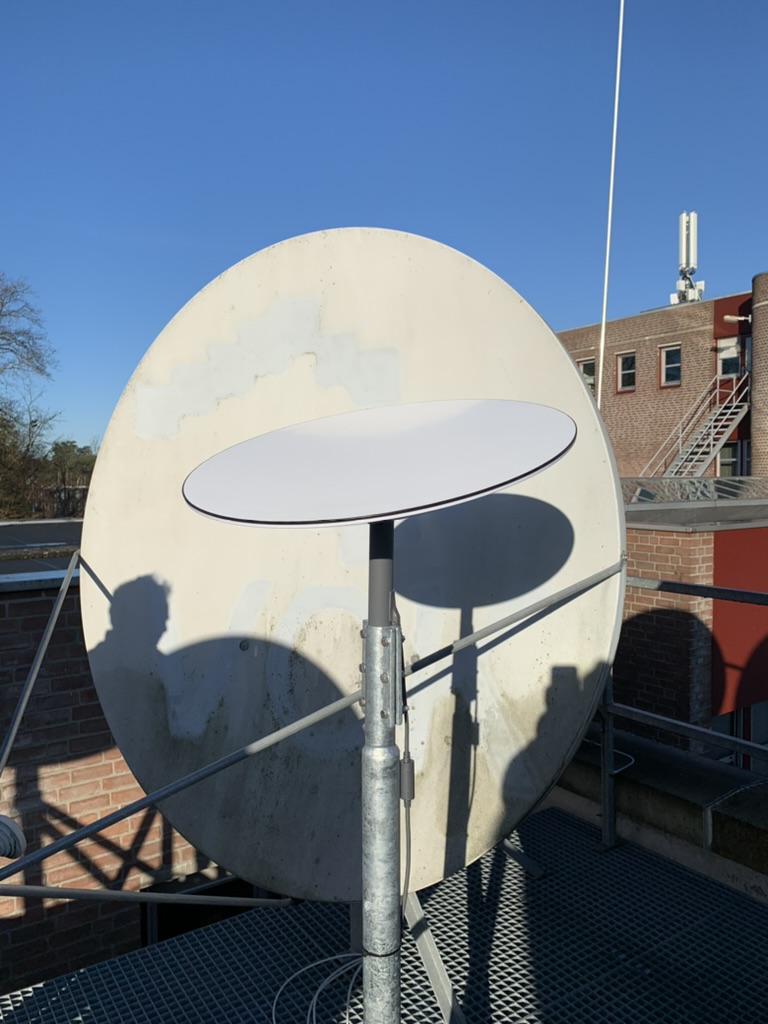}
        \label{fig:setup_Enschede}
    }
    \caption{Our measurement setups. In this picture, the Froggit DP2000 weather station can be seen placed directly next to the Starlink dish.}
    \label{fig:setup}
\end{figure}

We conduct Starlink measurements and collect weather data at two sites: (1) on the rooftop of one of the university buildings in Osnabrück, Germany (Fig.~\ref{fig:setup_Osnabrück}), and (2) on one of the university buildings in Enschede, the Netherlands (Fig.~\ref{fig:setup_Enschede}). The distance as the crow flies between both sites is 80\,km. This means it is likely that both sites connect to the same satellites in sequence. Furthermore, both measurement stations connect to the 53\degree\ orbit. At both positions, we deployed a similar measurement setup as shown in Fig.~\ref{fig:general_setup}.
Each setup consists of four devices: (1) a Starlink user terminal (a.k.a.\ a ``dishy''), (2) a Starlink router configured in pass-thru mode, (3) a measurement client on a VM running Ubuntu and (4) a Froggit~DP2000 weather station deployed directly next to the user terminal. The user terminals have an unobstructed view of the sky, and run the most up-to-date firmware as pushed to the terminal by Starlink. Both sites have a regular Starlink subscription. The measurement clients both have a secondary 1\,Gbit/s network interface on the local network, which we use to orchestrate measurements and to retrieve results. There are two key differences between both sites: the site in Osnabrück has a second generation user terminal (v2), whereas the site in Enschede has a first generation user terminal (v1). Secondly, the site in Enschede needs to contact the measurement server via a WAN interface; the measurements are instrumented such that any delays on this link do not impact results. We perform {\tt iperf} throughput measurements against a server located in Osnabrück. This server has a 5\,Gbit (shared) WAN connection to the Internet. % serves as endpoint for our {\tt iperf} measurements.

Based on publicly available information about Starlink ground stations~\cite{starlink-sx}, we assume that traffic from both locations will likely reach a ground station in Aerzen, Germany, which is 145\,km from Osnabrück and 167\,km from Enschede as the crow flies. From this ground station, we observe traffic entering the public Internet at Starlink's POP in Frankfurt.

Finally, we gather results on the data collection server, located in Osnabrück. This server records the following data:
%This server collects data from all link measurements, and it collects data from several more endpoints. The endpoints, the collector server is collecting, are the following:
\begin{itemize}[itemsep=1pt,topsep=1pt,labelindent=0.25em,leftmargin=*]
    \item throughput, RTT, packet loss and traceroutes;
    \item fine-grained weather data from the sensors on both sites;
    \item API data from public weather services for Germany (DWD) and the Netherlands (KNMI);
    \item all diagnostic data provided by the Starlink dishy (software version, obstruction, debug information, etc.).
\end{itemize}
%The collector is a self-developed python software, that collects the above-mentioned data and stores it into a \emph{MariaDB} database. Details on the collected values will be provided in the following sections.

\subsection{Network Link Measurement}
\begin{figure}
    \centering
    \includegraphics[width=0.7\columnwidth]{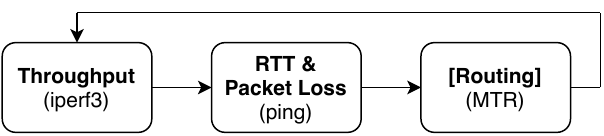}
    \caption{The measurement process of the network link parameters.}
    \label{fig:measurement_process}
\end{figure}

We use a three step sequential process to measure network link parameters, as shown in Fig.~\ref{fig:measurement_process}. The first two steps (the \textit{iperf3} and \textit{ping} measurements) are executed approximately every 3~minutes, the third step is executed approximately every 6~minutes.
%The process to measure the network link parameters is visualized in Fig.~\ref{fig:measurement_process}. 
%It is a three-step sequential process consisting of the throughput measurement of the up- and down-links, the parallel RTT and packet loss measurement, and an optional measurement of routing information.
%The throughput, RTT, and packet loss measurements are repeated every three minutes, whereas the routing measurement is conducted in an interval of six minutes.
%
%\subsubsection{Throughput}
We use \textit{iperf3} (version 3.9)~\cite{iperf} to measure up- and downstream throughput. 
The measurement of both directions is conducted in parallel using UDP with two separate measurements. While \textit{iperf3's} has a bidirectional mode, we do not use this mode since it is susceptible to CPU limitations which can lead to inaccurate measurement results.
%Furthermore, the bidirectional mode does not allow to set distinct target datarates for the up- and down-links.
%To avoid CPU-limitations and to save datarate, we conducted the parallel measurements by starting two separate \textit{iperf3} instances with target bitrates of 500~Mbit/s and 100~Mbit/s on the down- and up-link, respectively.
We limit the target bitrate for the uplink to 100\,Mbit/s and for the downlink to 500\,Mbit/s. This ensures \textit{iperf3} does not experience CPU limitations and that our measurements comply with Starlink's fair-use policy, ensuring that the datarate is not artificially limited by Starlink.
Each measurement takes 15 seconds with a five second timeout grace period.

%\subsubsection{Round Trip Time (RTT) and Packet Loss}
We utilize \textit{ping} to measure the round trip time (RTT) and packet loss. This measurement consists of a sequence of 250 packets with an interval of 0.1\,s. Aside from packet loss and average RTT we also report the best and worst RTT as well as the standard deviation.
%\footnote{\url{https://linux.die.net/man/8/ping}} utility to measure the round trip time (RTT) and packet loss. This measurement consists of a sequence of 250 packets with an interval of 0.1s. Besides the packet loss and average RTT, we also report the best and worst RTT, as well as the standard deviation.

%\subsubsection{Routing}\label{sec:setup_routing}
We collect routing information from our client to our server using \textit{Matt's traceroute} (MTR, version 0.94)~\cite{mtr} with 15~report cycles.
For each hop, the output includes the host IP or host name, packet loss information, and ping statistics. %, enabling a high traceability about the routing.

\subsection{Weather Measurement}
% RvRD: I reversed the order in which this is explained; a key highlight of your dataset
%       is the fact that you have a weather station right next to the dishy; you kind of
%       glossed over this major feature by first talking about DWD and KNMI data ;-)
%
To accurately capture weather conditions at our measurement positions, we place a Froggit DP~2000 weather station directly next to our Starlink dishes. Table~\ref{tab:froggit_features} lists the parameters the weather station collects. We also record calculated features produced by the weather station, specifically the dew point, windchill, wind gusts, and daily, weekly, monthly and annual rainfall. We also registered our weather stations with the Weather Underground~\cite{wunder_osn,wunder_ens}. Measurements are collected at 1-minute intervals. In addition to our own weather measurements, we also source data from the national weather services of Germany (DWD) and The Netherlands (KNMI). For Osnabrück, we collect data for station ID~00342 located at Belm, approximately 10.5\,km from our antenna, and for Enschede, we collect data for station ID~290 located at Twente Airport, approximately 4.5\,km from our antenna.

\begin{table}[t]
\centering
\caption{Measured Features of the Froggit DP~2000 weather station}
\resizebox{\linewidth}{!}{%
\begin{tabular}{lll}
\toprule
\textbf{Feature}                    & \textbf{Unit}    & \textbf{Accuracy}                                                                                           \\ \midrule
Temperature                         & \degree C        & $\pm$ 0.3\,\degree C                                                                                              \\
Humidity                            & \%               & $\pm$ 3.5\%                                                                                                    \\
Rain Volume                         & mm               & $\pm$ 10\%                                                                                  \\
Windspeed                           & m/s  & \begin{tabular}[c]{@{}l@{}}$<$ 10m/s: $\pm$ 0.5m/s \\ $\geq$ 10m/s: $\pm$ 5\%\end{tabular}           \\
Wind Direction                      & °                & \begin{tabular}[c]{@{}l@{}}$<$ 2m/s: $\pm$ 10\degree \\ $\geq$ 2m/s: $\pm$ 7\degree \end{tabular}                 \\
UV                                  & UV-Index         & Range: 0-15
                                                    \\
Barometric Pressure & hPa              & \begin{tabular}[c]{@{}l@{}}$\pm$ 5 hPa in 700 – 1,100 hPa range\\ resolution: 0.1~hPa (0.01~in Hg)\end{tabular} \\
Solarradiation                      & W/m\textsuperscript{2} & Not specified                                                                                  \\                                         \bottomrule
\end{tabular}
}
\label{tab:froggit_features}
\end{table}

%\begin{table}[t]
%\centering
%\caption{Calculated Features of the Froggit DP2000 weather station}
%%\resizebox{\linewidth}{!}{%
%\begin{tabular}{lll}
%\toprule
%\textbf{Feature}                    & \textbf{Unit}                                                                                              \\ \midrule
%Dew Point                           & °C                                                                                                         \\
%Windchill                           & °C                                                                                                         \\
%Wind Gust                           & m/s                                                                                            \\
%Rain Daily                          & mm                                                                                                         \\
%Rain Weekly                         & mm                                                                                                         \\
%Rain Monthly                        & mm                                                                                                         \\
%Rain Yearly                         & mm                                                                                                         \\
%
%\bottomrule
%\end{tabular}
%%}
%\label{tab:froggit_calculated_features}
%\end{table}

\subsection{Dataset}

Based on the measurement architecture discussed above, we perform measurements resulting in the WetLinks dataset, which we release publicly together with all pre-processing scripts~\cite{wetlinks-data}. Our measurements in Osnabrück start from September~14\textsuperscript{th}, 2023, and our measurements in Enschede start on October~12\textsuperscript{th} of the same year. Both measurements are still running at the time of writing and will run until at least the end of March~2024. This results in a dataset of continuous measurements spanning 6.5~months for Osnabrück, with approximately 80,000 data points, and 5.5~months for Enschede, with approximately 60,000 data points.

We use a simple comma-separated format (CSV) for data. The raw dataset consists of CSV files that respectively record throughput measurements ({\tt net\_iperf.csv}), latency measurements ({\tt net\_ping.csv}), traceroute measurements ({\tt net\_traceroute.csv}), debug data from the Starlink dishes ({\tt starlink.csv}), and finally data from the weather stations positioned next to the dishes ({\tt froggit.csv}). The pre-processing scripts we provide can be used to combine these datasets and harmonize the different timescales at which the measurement data is collected (e.g., averaging weather data, which is collected at higher frequency than throughput measurements). Providing the raw data together with the pre-processing scripts allows users of the data to fully reproduce results and to make their own choices in harmonizing data for the purpose for which they wish to use the data.

Finally, we note that the dataset comes with a few limitations. First, our throughput measurements occasionally fail. This leads to empty fields in the CSV file that can easily be filtered out. Second, our traceroute measurements also occasionally fail, likely due to intermittent Starlink outages. This leads to traceroutes that contain only two hops and then time out. Finally, our weather station in Enschede suffered a two-week outage due to the battery running out over the end-of-year break; this means weather data for this period is missing for the Enschede location.

%\input{Dataset}
% !TEX root = main_ieee.tex

\section{Analysis}\label{sec:analysis}

In this section, we discuss the analyses we performed on the WetLinks dataset. Our focus is on verifying the quality of our dataset. We do this by comparing our analysis to earlier studies for different metrics. %We analyse four aspects: throughput, packet loss, latency and the influence of weather conditions on throughput.

%We now analyze our dataset. First, we conduct a general analysis of the Starlink network performance in terms of the throughput, packet loss, latency, and routing in Sec.~\ref{sec:analysis_throughput}, Sec.~\ref{sec:analysis_loss}, 
%Sec.~\ref{sec:analysis_latency}, and Sec.~\ref{sec:analysis:routing}, respectively.
%Then, we analyze the impact of the daytime on the network performance in Sec.~\ref{sec:analysis_daytime}. Finally, we analyze the impact of different weather conditions on the network performance in Sec.~\ref{sec:analysis:weather}.

\begin{figure}
    \centering
    \subfloat[Download Throughput]{
        \includegraphics[width=0.4\linewidth]{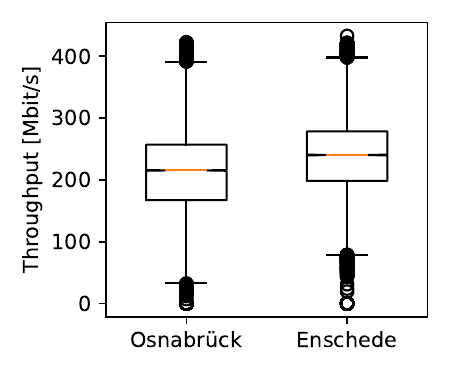}
        \label{fig:dl_overall}
    }
    \subfloat[Upload Throughput]{
        \includegraphics[width=0.4\linewidth]{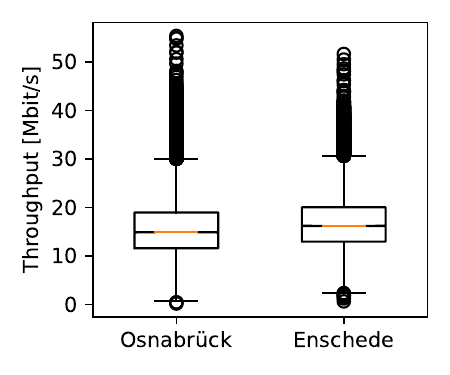}
        \label{fig:ul_overall}
    }
    \caption{Download and upload throughput for Osnabr\"{u}ck and Enschede.}
    \label{fig:dl_ul_overall}
\end{figure}

\subsection{Throughput Analysis}\label{sec:analysis_throughput}
Fig.~\ref{fig:dl_ul_overall} shows boxplots of download and upload throughput.
% RvRD: since the table shows exactly the same information as the boxplots,
% I took the liberty of commenting it out to save space.
% Statistics, including the mean, median, and 25th and 75th percentiles can be found in Tab.~\ref{tab:throughput}.
%
The large inter-quartile ranges and whiskers indicate a large variability in performance for both down- and upload. The download measurements range from 0\,Mbit/s, representing short and rare Starlink outages, to more than 400\,Mbit/s, significantly exceeding the advertised maximum download throughput of 220\,Mbit/s. 
The upload measurements range from 0\,Mbit/s to more than 50\,Mbit/s.
The speeds measured in Enschede consistently exceed the ones in Osnabr\"{u}ck. % in both, download and upload throughput. 
Considering that both are connected to the 53\degree\ orbit and their geographic proximity, we speculate that this may be caused by the difference in dish versions.

Comparing our results to existing studies is challenging, as these often use different transport protocols that are sensitive to packet loss (e.g., TCP or QUIC). % are used, which are sensitive to packet loss. 
Michel et al.\ \cite{michel2022first} report TCP download throughputs of up to 400\,Mbit/s, comparable to our UDP measurements.
Furthermore, they reported median TCP download and upload throughputs of 178\,Mbit/s and 17\,Mbit/s, respectively. In contrast, Kassem et al.\ \cite{kassem2022browser} report median throughputs of 123\,Mbit/s and 11\,Mbit/s. 
Our measured download throughputs are significantly higher, highly likely due to the use of loss-sensitive congestion control algorithms, e.g., TCP CUBIC, in \cite{kassem2022browser} and \cite{michel2022first}.

%\begin{table}[t]
%\centering
%\caption{Throughput statistics of our measurements in Mbit/s}
%\resizebox{\linewidth}{!}{%
%\begin{tabular}{lllll}
%\toprule
%\textbf{Location} & \textbf{Mean}                    & \textbf{Median} & \textbf{25th-percentile} & \textbf{75th-percentile}                                                                                              \\ \midrule
%\multicolumn{2}{l}{\textbf{Download Throughput}} \\
%Osnabrück                           & 213 & 216.1 & 169.6 & 257.4\\
%Enschede & 232.1 & 234.8 & 194.7 & 270.9 \\
%\multicolumn{2}{l}{\textbf{Upload Throughput}} \\
%Osnabrück & 15.9 & 14.6 & 11.2 & 19 \\
%Enschede & 16.7 & 15.9 & 12.6 & 19.8 \\
%\bottomrule
%\end{tabular}
%}
%\label{tab:throughput}
%\end{table}

\subsection{Packet Loss Analysis}\label{sec:analysis_loss}

We next turn our attention to packet loss. Packet loss occurs frequently during our measurements. The overall statistics in Tab.~\ref{tab:plr_statistics} show that well over 25\% of our measurements contain some packet loss. The minimum packet loss rate (PLR) was 0.4\%, which corresponds to one out of 250 packets sent during the loss measurement is lost. 
%
% RvRD while the explanation of the 0.4% is useful, the stats below just repeat what is shown in the table.
% The highest observed PLRs in Osnabrück and Enschede were 84.4\% and 81.2\%, respectively.
%In Osnabrück, the measurements, in which packet loss occurred, have a mean PLR of 1.03\%. In Enschede, the mean PLR is 1.0\%.

\begin{figure}
    \centering
    \includegraphics[width=0.9\linewidth]{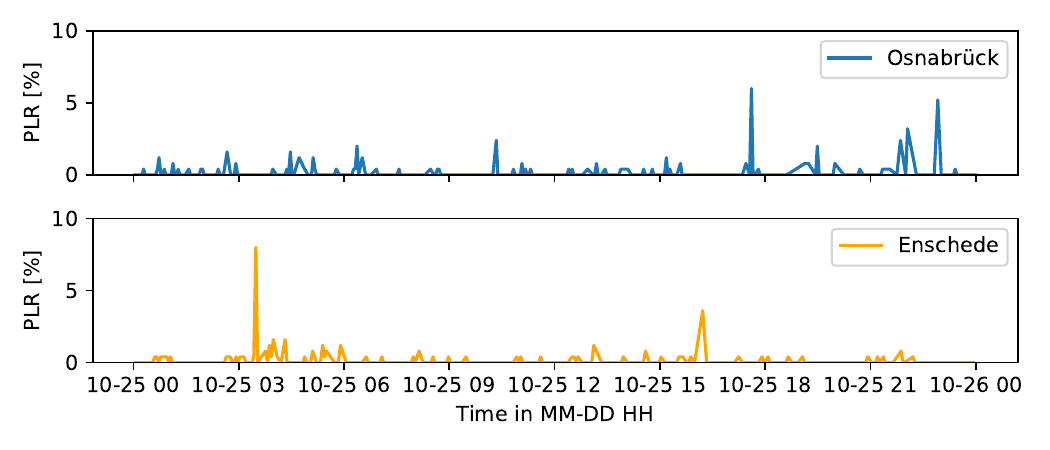}
    \caption{The PLR plotted for the 25th of October 2023 from our measurements.}
    \label{fig:plr_day}
\end{figure}

\begin{figure} 
    \centering
    \includegraphics[width=0.8\linewidth]{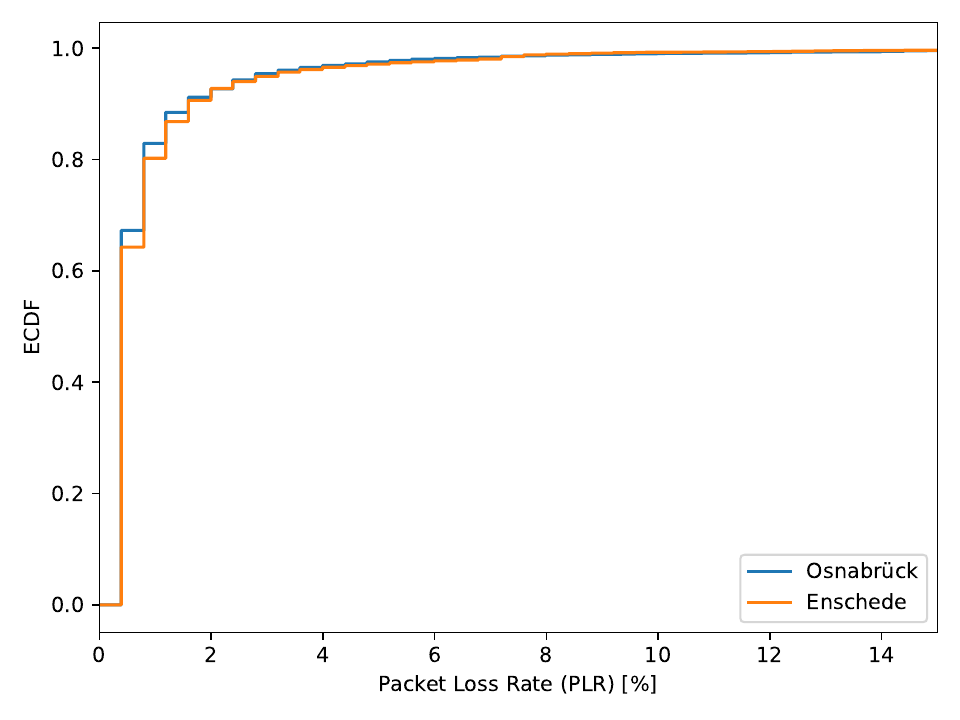}
    \caption{ECDFs of the packet loss rates.}
    \label{fig:packetloss_ecdfs}
\end{figure}

Fig.~\ref{fig:plr_day} shows the packet loss pattern over a single randomly selected 24-hour period. While the PLR is low most of the time, short spikes can be observed. These spikes typically only affect single measurements and can be interpreted as short burst losses. There is no evidence to suggest a correlation between loss events in Osnabrück and Enschede. Fig.~\ref{fig:packetloss_ecdfs} shows an ECDF of the PLR over all our measurement data split between the two sites. As the plot shows, the vast majority ($\pm$65\%) of loss events concern only a single packet (PLR 0.4\%). Furthermore, over 80\% of measurements have a PLR $\leq$1\%, and only some 3\% of measurements have a PLR $\geq$5\%.

Related work by Ma et al.~\cite{ma2023network} reports cyclic burst loss patterns over 12~hours. We do not see similar patterns in our data.
Michel et al.~\cite{michel2022first} report a PLR of 0.4\% on the downlink and 0.45\% on the uplink for low-load periods. Our data shows comparable behaviour with additional short spikes caused by burst losses.
These burst losses have also been reported by Kassem et al.~\cite{kassem2022browser} who found that they likely primarily occur during satellite handovers.

% RvRD: edited the table to make it more readable in a single column
\begin{table}[t]
\centering
\caption{Packet Loss Statistics}
\resizebox{\linewidth}{!}{%
\begin{tabular}{lrrrrr}
\toprule
\textbf{Site}   
    & \textbf{Samples with loss}  
    & \textbf{Min.\ PLR} 
    & \textbf{Max.\ PLR} 
    & \textbf{Mean PLR} 
    & \textbf{$\sigma$ PLR}                                                                                   
    \\ \midrule
% RvRD: I made the number of significant decimals consistent
Osnabrück   & 28.4\% %22.72    
            & 0.4\% %0.4
            & 87.2\% %84.4
            & 1.0\% %1.03
            & 2.8\% \\ %3.27 \\
Enschede    & 33.4\% %24.6
            & 0.4\% %0.4
            & 84.0\% %81.2
            & 1.0\% %1.0
            & 2.4\% \\ %2.88\\ \bottomrule
\end{tabular}
}
\label{tab:plr_statistics}
\end{table}

\subsection{Latency Analysis}\label{sec:analysis_latency}

\begin{figure}
    \centering
    \subfloat[The overall RTT]{
        \includegraphics[width=0.4\linewidth]{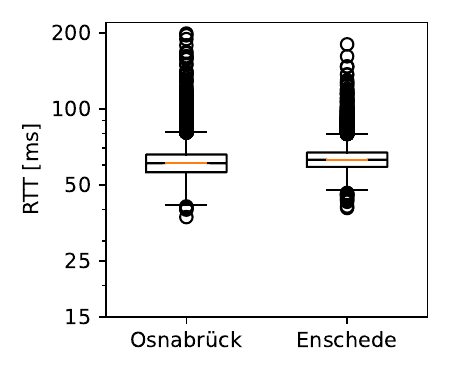}
        \label{fig:rtt_full}
    }
    \subfloat[The bent-pipe latency]{
        \includegraphics[width=0.4\linewidth]{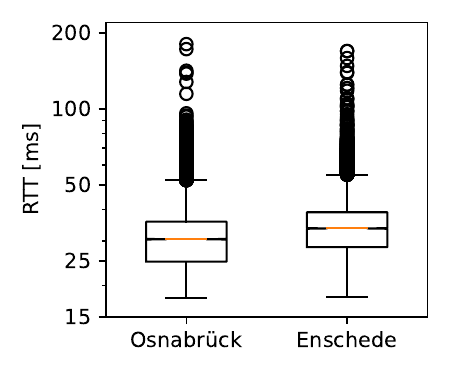}
        \label{fig:bent-pipe_latency}
    }
    \caption{Boxplots of the latency.}
    \label{fig:latency_overall}
\end{figure}

Fig.~\ref{fig:rtt_full} shows the overall RTT distribution for the route from the measurement node to our server. The mean RTTs are 61.52\,ms ($\sigma$=7.81\,ms) and 63.53\,ms ($\sigma$=6.87\,ms) for Osnabrück and Enschede respectively. Approximately 72\% of measurements have an RTT within 1$\sigma$, with outliers up to 200\,ms.

%For our measurements from Osnabrück, the mean RTT is 62.93ms with a standard deviation ($\sigma$) of 7.66ms. 71.63\% of the RTT measurements are within one $\sigma$. For our measurements from Enschede, the mean RTT is 64.49ms with a $\sigma$ of 6.36ms. 73.27\% of the measurements are within one $\sigma$.
%In the boxplots, outliers up to 200ms can be observed.

To further analyze the cause of the high variance, we examine the RTT distribution for the bent-pipe section of the route in Fig.~\ref{fig:bent-pipe_latency} based on the RTT data from our traceroute measurements. The bent-pipe has a mean RTT of 31.08\,ms ($\sigma$=11.18\,ms) for our location in Osnabrück and 34.56\,ms ($\sigma$=12.0\,ms) for Enschede. Approximately 85\% of measurements have an RTT within 1$\sigma$, with outliers up to 200\,ms. These numbers indicate that RTT variance of the complete link is mostly caused by the bent-pipe section of the path.

%To better analyze the cause of the high variance, we plotted only the RTT measurements from the bent-pipe hop of our traceroute measurements in Fig.~\ref{fig:bent-pipe_latency}. Note that these RTT measurements were conducted using \textit{mtr} with only 15 packets sent per measurement (cf. Sec.~\ref{sec:setup_routing}), making the results more prone to outliers compared to our overall RTT measurements. 
%The bent-pipe has a mean RTT of 32.63ms with a $\sigma$ of 9.11ms for the measurements from Osnabrück. Furthermore, 75.38\% of those measurements are within one $\sigma$. The measurements from Enschede had a mean bent-pipe RTT of 35.45ms with a $\sigma$ of 16.6ms. 95.93\% of the measurements were within one  $\sigma$. In the boxplots of Fig.~\ref{fig:bent-pipe_latency}, outliers of up to 200ms can be observed for both measurement sites. Additionally, one outlier of 554ms occurred in Osnabrück, and one outlier of 1389ms occurred in Enschede.
%Large parts of the standard deviations are caused by these two outliers. Removing them leads to a mean RTT of 32.06ms, median RTT of 32.3ms, a $\sigma$ of 8.18ms, and 70.61\% of the measurements within one $\sigma$ for Osnabrück, and a mean RTT of 35.31ms, a median RTT of 34.41ms, a $\sigma$ of 8.85ms, and 73.5\% of the measurements within one $\sigma$ for Enschede.
%These numbers indicate that RTT variance of the complete link is mostly caused by the bent-pipe.

Overall, the average bent-pipe RTT of 31-35\,ms is well within the expected range from previous works \cite{mohan2023multifaceted,kassem2022browser} of 30-40~ms for measurements in the EU on the 53\degree\ orbit.
Our conclusion that the variance in latency is primarily caused by the bent-pipe section of the path also aligns with previous work~\cite{izhikevich2023democratizing,kassem2022browser,ma2023network,michel2022first}.

% Furthermore, our data suggests that the RTT of a Starlink link is prone to high variance primarily caused by the bent-pipe. This is also consistent with previous work \cite{izhikevich2023democratizing}, \cite{kassem2022browser}, \cite{ma2023network}, \cite{michel2022first}. In the past, these fluctuations were primarily explained with sub-optimal satellite handovers \cite{raman2023dissecting}, constant satellite movements \cite{ma2023network}, or Starlink's 15 second reconfiguration interval \cite{mohan2023multifaceted}.
%
% RvRD: I suggest leaving the sentences below out of the paper; they don't contribute new information,
% just speculation that needs further work. If you want to say this, it belongs in future work (but I think
% it's a bit too detailed for that).
%While single extreme outliers might be caused by sub-optimal satellite handovers, our observed relatively strong constant ground-level fluctuations occur in a frequency that is too high to be solely caused by them. The ground-level fluctuations are likely caused by a combination of the satellite movements and Starlink's 15 seconds reconfiguration interval. To validate this, a continuous RTT measurement over a longer time-period alongside a precise tracking of the satellite positions mapped to the measurement timestamps would be necessary.

\subsection{Impact of The Time of Day}\label{sec:analysis_daytime}

\begin{figure*}
\centering
    \subfloat[Download Throughput Osnabrück]{
        \includegraphics[width=0.49\linewidth]{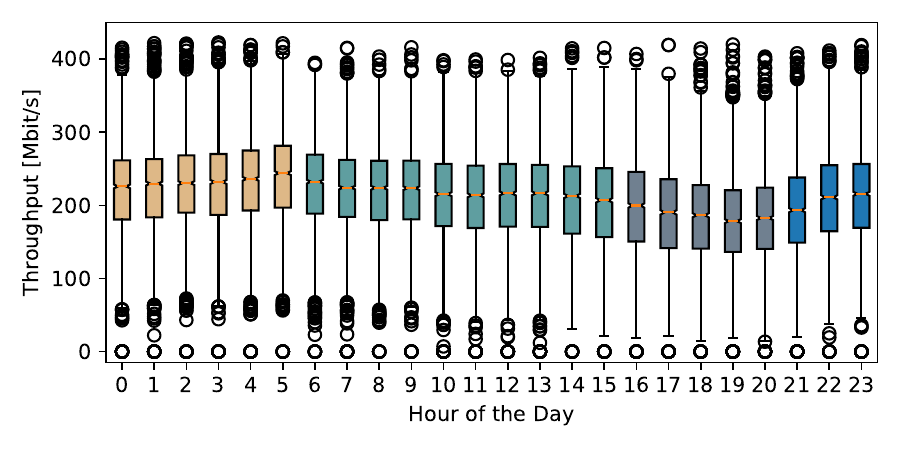}
        \label{fig:day_dl_uos}
    }
    \subfloat[Download Throughput Enschede]{
        \includegraphics[width=0.49\linewidth]{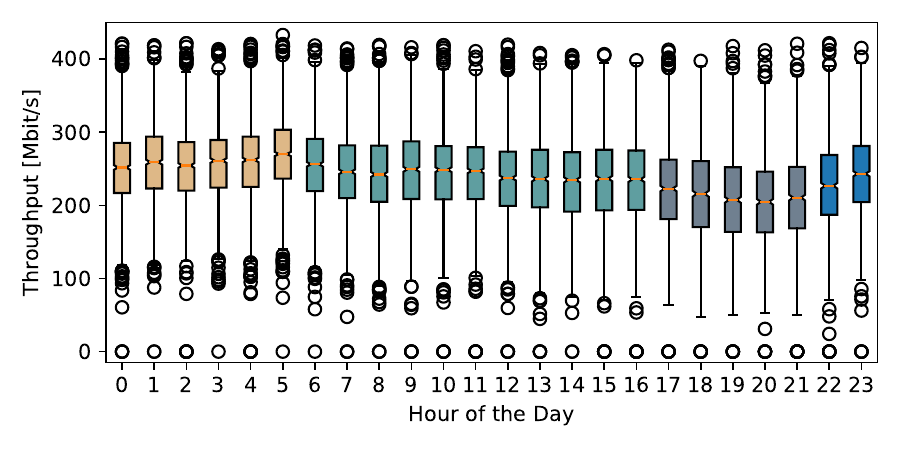}
        \label{fig:day_dl_twente}
    }
    
    \vspace{-1em}

    \subfloat[Upload Throughput Osnabrück]{
        \includegraphics[width=0.49\linewidth]{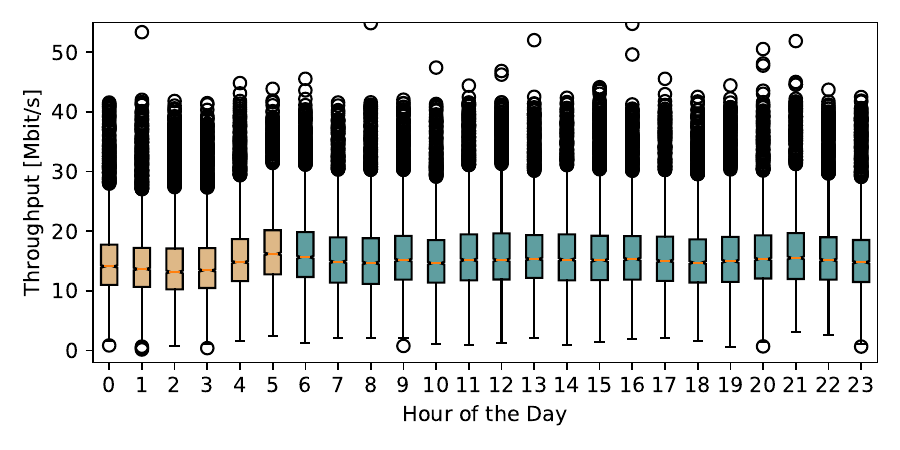}
        \label{fig:day_ul_uos}
    }
    \subfloat[Upload Throughput Enschede]{
        \includegraphics[width=0.49\linewidth]{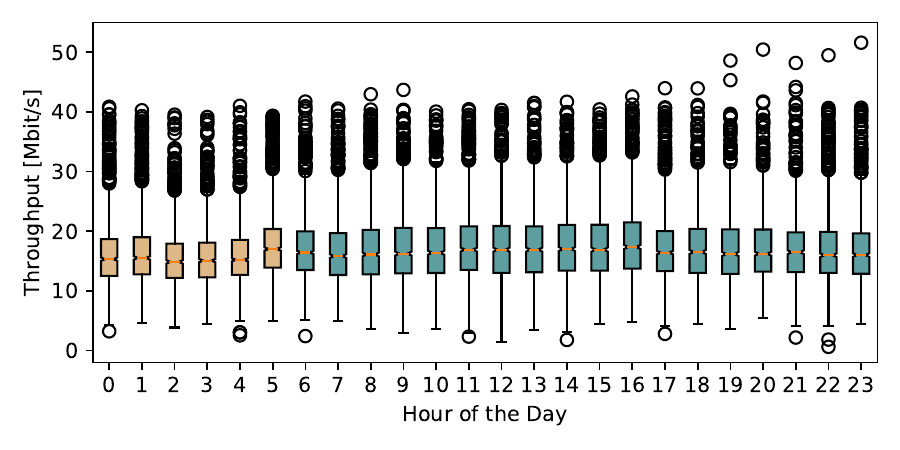}
        \label{fig:day_ul_twente}
    }
    
%    \vspace{-1em}

%    \subfloat[RTT Osnabrück]{
%        \includegraphics[width=0.49\linewidth]{Plots/day_rtt_uos.pdf}
%        \label{fig:day_rtt_uos}
%    }
%    \subfloat[RTT Enschede]{
%        \includegraphics[width=0.49\linewidth]{Plots/day_rtt_twente.pdf}
%        \label{fig:day_rtt_twente}
%    }
    
%    \vspace{-1em}

%    \subfloat[Packet Loss Osnabrück]{
%        \includegraphics[width=0.49\linewidth]{Plots/day_loss_uos.pdf}
%        \label{fig:day_loss_uos}
%    }
%    \subfloat[Packet Loss Enschede]{
%        \includegraphics[width=0.49\linewidth]{Plots/day_loss_twente.pdf}
%        \label{fig:day_loss_twente}
%    }
    
    \caption{All measurements binned into 24 bins representing the hours of a day. The different periods are colored.} %Measurements in Osnabrück started about one month earlier than in Enschede, making the dataset significantly larger. This leads, for example, to more outliers and makes an exact numerical comparison difficult.}
    \label{fig:day}
\end{figure*}

%Coming up, we analyze the impact of the time of day on the network parameters, visualized in Fig.~\ref{fig:day}. Generally, similar behaviors can be observed for both measurement sites.
Fig.~\ref{fig:day} shows the impact of the time of day on network parameters. Below, we analyze the impact of the time of day in detail for the different network parameters.
%Coming up, we analyze the impact of the time of day on the network parameters, visualized in Fig.~\ref{fig:day}. Generally, similar behaviors can be observed for both measurement sites.

\textbf{Download Throughput:}
% RvRD (20240219): I suggest shading the background of the plots with alternating colours in
% a zebra-stripe pattern to show the different periods.
Fig.~\ref{fig:day_dl_uos} and Fig.~\ref{fig:day_dl_twente} illustrate that a typical day can be split into multiple periods. 
In the early morning, between 0h-5h, the median download throughput is around~230\,Mbit/s in Osnabrück and 250\,Mbit/s in Enschede. From 6h to approximately 15h, it decreases to around~210\,Mbit/s in Osnabrück and 235\,Mbit/s in Enschede. Then, a further drop can be observed until 19h, which sees the lowest throughput of 180\,Mbit/s and 205\,Mbit/s in Osnabrück and Enschede, respectively. Afterwards, the throughput recovers, reaching 210\,Mbit/s in Osnabrück and 225\,Mbit/s in Enschede around 22h.

Our findings are consistent with previous measurements within the EU on the 53\degree\ orbit \cite{garcia2023multi,kassem2022browser}, where the maximum download throughput was found around 5h \cite{garcia2023multi}, or between 0h-6h \cite{kassem2022browser}.
Additionally, Kassem et al.~\cite{kassem2022browser} reported that the maximum throughput can be twice as high as the minimum throughput on a day. While this might be the case for single days, our data suggests that on average, the minimum download throughput of a day is approximately 20\% lower than the maximum. This is also consistent with findings by Michel et al.~\cite{michel2022first}, who report fluctuations around $\pm10\%$.
%We note that a detailed comparison to related work based on throughput is difficult, because it was shown in \cite{kassem2022browser}, that the throughput can significantly differ between different regions of the world. However, the relevant measurements we refer to in \cite{garcia2023multi,kassem2022browser,michel2022first}, were all conducted in European countries, such as our measurements, assuming a sufficient comparability. %  - Karlstad, Sweden in \cite{garcia2023multi}, London, United Kingdom in \cite{kassem2022browser}, and Louvain-la-Neuve, Belgium in \cite{michel2022first}. 
%Because of the physical proximity of our measurement places, we assume a sufficient comparability to be able to make the aforementioned relations.

Our data does not provide a clear explanation for the observed fluctuations. Since the routes remain largely constant also during the hours with lower throughput, it is likely that the general load within the Starlink network causes the observed behavior, especially in the evening hours from 17h-22h.% We suspect a higher load during the evening hours between 17:00h and 22:00h, possibly leading to load management inside the Starlink network for standard, non-priority traffic.

\textbf{Upload Throughput:}
Fig.~\ref{fig:day_ul_uos} and \ref{fig:day_ul_twente} indicate an increasing upload throughput between 0h-5h, and an almost constant one for the rest of the day. The minimum throughput is reached around 2h with 13.19\,Mbit/s in Osnabrück, and 14.87\,Mbit/s in Enschede. The maximum throughput is reached at 5h and with 16.23\,Mbit/s in Osnabrück, and at 16h with 17.37\,Mbit/s in Enschede. 
The interquartile ranges, the whiskers and the outliers indicate a strongly fluctuating upload throughput between 0\,Mbit/s and 40\,Mbit/s.
Interestingly, there are almost no Starlink outages at 0\,Mbit/s. This implies that the Starlink uplink appears to be more stable than the downlink.
Similar to the download throughput our data shows similar behaviour to previous studies \cite{kassem2022browser}.

\textbf{RTT and Packet Loss:}
Neither the RTT, nor the packet loss are impacted by the time of day. The median RTT is 62\,ms $\pm$ 3\,ms throughout the day. The median PLR is constantly 0.4\%, with occasional burst losses that can occur at any time.

%The median RTT (cf. Fig.~\ref{fig:day_rtt_uos}, Fig.~\ref{fig:day_rtt_twente})is relatively constant throughout the day. It fluctuates around 63ms, with $\pm$3ms. The whiskers and large outliers up to 200ms indicate that similarly to the download and upload throughputs, also the RTT is prone to strong fluctuations.

%\textbf{Packet Loss:}
%Fig.~\ref{fig:day_loss_uos} and Fig~\ref{fig:day_loss_twente} show boxplots of only the measurements in which loss occurred. While the median PLR is always 0.4\%, it is 0.8\% at 04:00h in Osnabrück and at 04:00h and 05:00h in Enschede. The outliers indicate rare spikes in the PLR, which can be interpreted as short burst losses and which can happen at any time. There appears to be no time-based component in the loss.

\subsection{Weather Impact}\label{sec:analysis:weather}

\begin{figure*}
    \centering
    \subfloat[Osnabrück]{
        \includegraphics[width=0.4\linewidth]{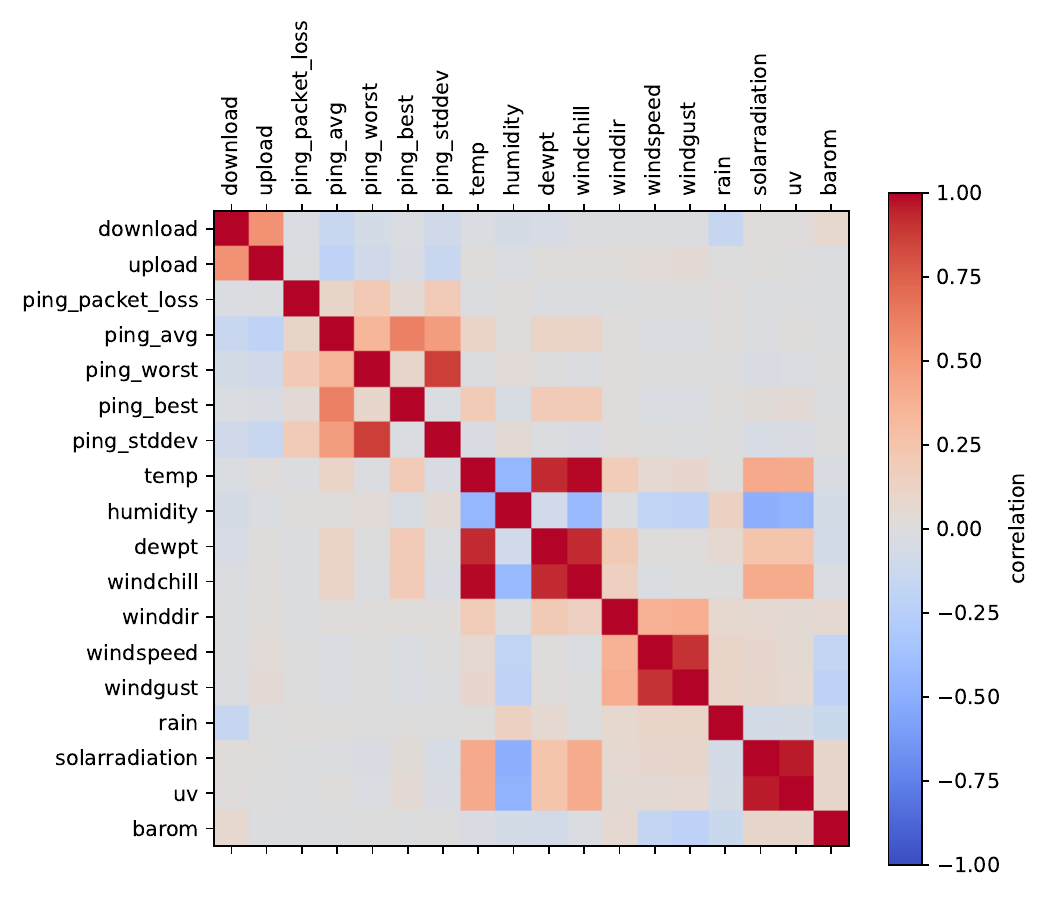}
        \label{fig:weather_corr_os}
    }
    \subfloat[Enschede]{
        \includegraphics[width=0.4\linewidth]{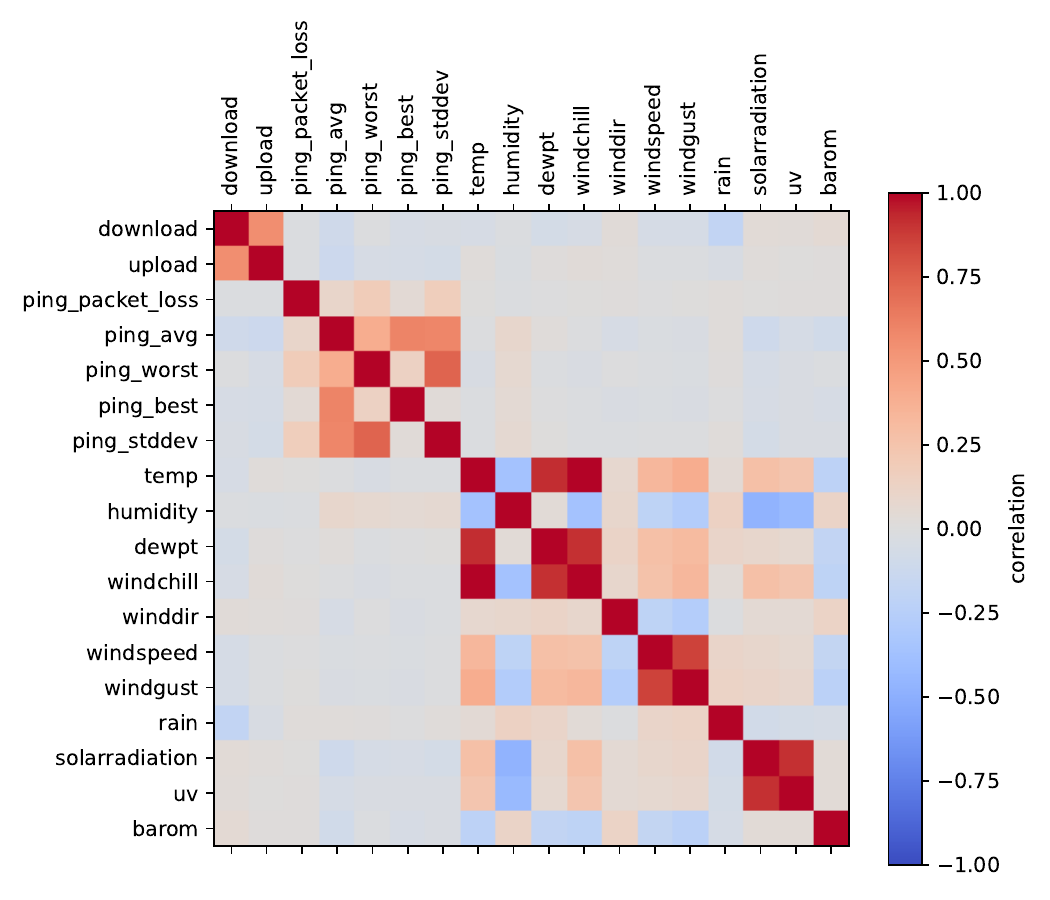}
        \label{fig:weather_corr_enschede}
    }
    \caption{Weather Correlation Matrix}
    \label{fig:weather_corr_matrix}
\end{figure*}

\begin{figure}
    \centering
    \includegraphics[width=0.8\linewidth]{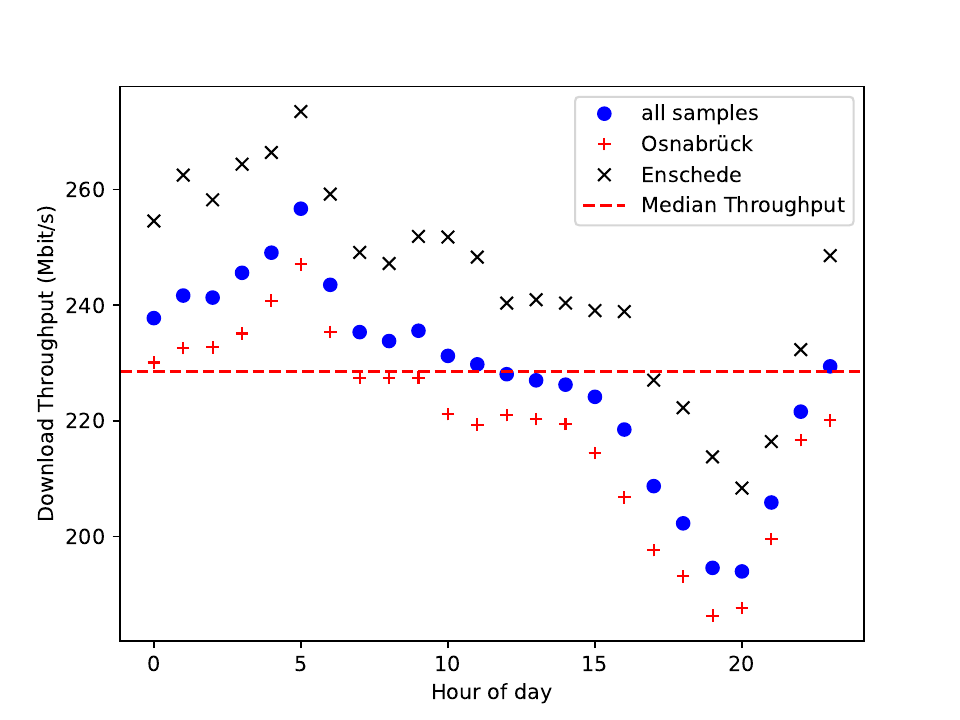}
    \caption{Median download throughput rate per hour}
    \label{fig:norain_median}
\end{figure}

\begin{figure*}
    \centering
    \subfloat[Osnabrück]{
        \includegraphics[width=0.4\linewidth]{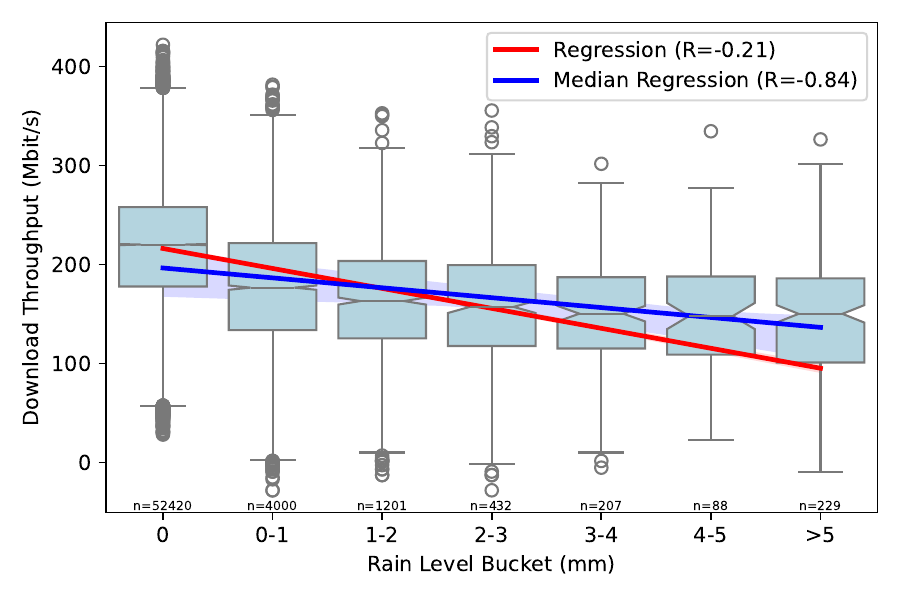}
        \label{fig:uos_rain_dl}
    }
    \subfloat[Enschede]{
        \includegraphics[width=0.4\linewidth]{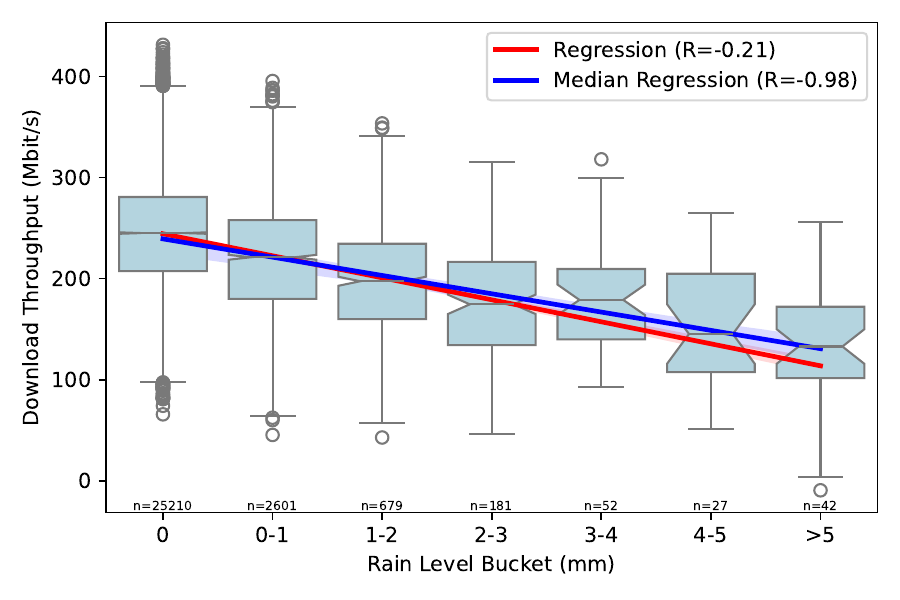}
        \label{fig:ut_rain_dl}
    }
    \caption{Download throughput grouped by rain level buckets, with regression functions. The median regression is computed on the median values for each bucket, to remove the noise and variance, caused by unknown effects.}
    \label{fig:rain_dl}
\end{figure*}
\begin{table}[]
\caption{Download throughput correction factors $c_{dt}$ to eliminate the time of day influence}
\label{tab:tod_correction}
\centering
\begin{tabular}{@{}c|r||c|r@{}}
\toprule
\textbf{Hour} & \textbf{$c_{dt}$} & \textbf{Hour} & \textbf{$c_{dt}$} \\ \midrule
0             & -9.24        & 12   & 0.46  \\
1             & -13.14       & 13   & 1.51  \\
2             & -12.79       & 14   & 2.26  \\
3             & -17.07       & 15   & 4.36  \\
4             & -20.55       & 16   & 10.02 \\
5             & -28.15       & 17   & 19.80 \\
6             & -14.99       & 18   & 26.23 \\
7             & -6.82        & 19   & 33.93 \\
8             & -5.27        & 20   & 34.55 \\
9             & -7.06        & 21   & 22.63 \\
10            & -2.71        & 22   & 6.93  \\
11            & -1.25        & 23   & -0.90 \\ \bottomrule
\bottomrule
\end{tabular}%
\end{table}

As discussed earlier in Section~\ref{sec:related_work}, weather conditions can have an impact on satellite communication. We created a correlation matrix shown in Fig.~\ref{fig:weather_corr_matrix}
to analyze their impact on the Starlink network parameters. Rain shows a weak negative correlation of about -0.165 and -0.192 to the download throughput for the measurements in Osnabrück and Enschede, respectively. All other weather factors have a correlation coefficient to the network parameters of close to zero, indicating no correlation.

\subsubsection{The Impact of Rain}
We continue our analysis with the hypothesis that rain intensity has an impact on download throughput. To verify this hypothesis, we first perform preprocessing on the data. We start by removing outliers using the interquartile range (IQR) with factor $1.5$. Next, we perform a preprocessing step to remove the time of day impact on the download throughput, as shown in Fig.~\ref{fig:day_dl_uos} and~\ref{fig:day_dl_twente}. For that, we grouped our data into hourly buckets and performed a median correction for each bucket. Specifically, we calculated the download throughput median of all non-rainy samples of the complete dataset, and the median of the non-rainy samples within each hourly bucket. 
%Which was done by calculating the median for all samples without rain of both datasets, grouped by the hour of day. Next, we calculated the overall median across all non-rainy samples for all samples of both datasets. 
The result is shown in Fig.~\ref{fig:norain_median}. 
%We can clearly observe the influence of the time of day in this plot, which is on an interval of $60$ Mbit/s at max. 
We can clearly observe the influence of the time of day, which is consistent at both measurement sites (same timezone), with a difference of over 60\,Mbit/s between the lowest and highest throughput. 
Using this data, we calculate the difference of the hourly median to the overall median and apply this difference as a correction factor ($c_{dt}$), as listed in Table~\ref{tab:tod_correction}, to corresponding samples. We used a product of both datasets for this step to come up with a more robust correction factor, by having more non-rainy samples. 
%
% RvRD: check if you agree with this description of the limitations
A limitation of this approach is that it does not account for inter-day variance (e.g., weekday/weekend, \ldots).

\begin{figure}
    \centering
    \includegraphics[width=0.8\linewidth]{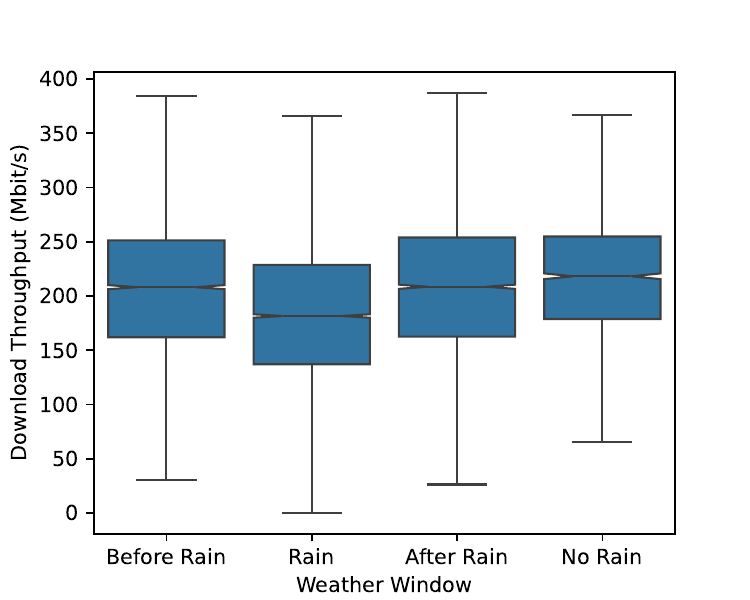}
    \caption{Boxplots for different weather conditions. For each rain window, 4 samples before and after the rain are used. We assume cloudy conditions before and after rain. In the no rain case, only samples with high solar radiation are used, to ensure that there are no clouds.}
    \label{fig:cloudanalysis}
\end{figure}

% RvRD: over which period do you count the level of precipitation? Per hour? Make this explicit
To then verify the hypothesis of rain influence, we group the samples into rain buckets. Each bucket contains all samples with $r>u$ and $r\leq v$, where $r$ is the level of precipitation in mm, $u$ the lower threshold, and $v$ the upper threshold of the bucket. We created 7 buckets, while the last bucket contains all samples with a precipitation $>$5\,mm. Additionally, we calculated a regression line, to test if an increasing rain level correlates with a change in download throughput. As the plots in Fig.~\ref{fig:rain_dl} show, there is a clearly visible correlation, having an R value of -0.21 at both sites, indicating a weak correlation. A null hypothesis significance test showed a $p$-value below 0.05, which confirms our hypothesis that rain influences the download throughput significantly. 
The boxplots show large whiskers, indicating that factors other than rain and the time of day have a significant impact on the download throughput. These large whiskers are the main cause of the weak correlation. To reduce this noise and to further isolate the impact of rain, we added regressions through the medians of the buckets, which have an R of -0.84 and -0.98, indicating a strong linear correlation between rain and download throughput. We argue that this is a valid approach for noise reduction since the confidence intervals of the medians as well as the IQRs are small. 
We also tested the upload throughput against rain buckets, however, having $R$ values below 0.05 for both locations, we do not observe a significant correlation here. 
This may be explained by the lower bandwidth offered for upload, which provides more room for adaptation and mitigation strategies. 

\subsubsection{The Impact of Clouds}
Clouds typically coincide together with rain and may also affect the download throughput. To differentiate the impact of clouds from the impact of rain, we looked at all rain periods in our dataset and extracted four samples before and after each rain window. These four samples translate to a time window of approximately 12~minutes. We assume that there is a cloudy condition shortly before and after rain periods. For comparison, we created a \emph{No Rain} group, which contained all samples without rain and solar radiation higher than 300\,W/m\textsuperscript{2}, to ensure that there are no or only a few thin clouds. Then, we compared these four groups, as visualized in Fig.~\ref{fig:cloudanalysis}. When analyzing these results, we observe that the worst throughput occurs during rain with a median of 181.52\,Mbit/s (95\%-CI: 180.09, 182.96), which is around 17\% lower than the median of the no rain condition group at 218.12\,Mbit/s (95\%-CI: 215.60, 220.64). Looking at the 4~sample windows before and after the rain period, we observe median throughputs of 208.08\,Mbit/s before the rain (95\%-CI: 206.36, 209.80) and 208.47\,Mbit/s after the rain (95\%-CI: 206.73, 210.21), which is almost 10\,Mbit/s lower than the median of the no rain condition group. These findings support our hypothesis that a cloudy sky also has a negative impact on the download throughput rate. However, there is a stronger negative impact of rain. To further investigate this issue, more research with an accurate labeling of the cloud situation is needed. We plan to extend our measurements in future work with this feature.

Compared to related work, our result is similar to Kassem et al.'s findings~\cite{kassem2022browser}. The authors observed an increase in page load time during rainy conditions. 
This also holds true for Ma et al.~\cite{ma2023network}, who discuss the impact on upload and whose study strongly suggested that mainly the download throughput is affected by rain. A drop of around 45\% in UDP download throughput for 4.1-5.2~mm rain was reported. 
Based on our medians for the buckets with 4-5~mm rain, we observed comparable drops of 31\% and 30\% for the measurements from Osnabr\"{u}ck and Enschede, respectively.
Ma et al.~\cite{ma2023network} also indicated a correlation with temperature. Our correlation matrix, however, does not indicate such a correlation. 
%
% RvRD: your data already includes winter data, so maybe this should be removed.
% We plan to analyze this issue further, when our dataset will be extended with measurements from the winter period.

% !TEX root = main_ieee.tex

\section{Conclusion}\label{sec:conclusion}

In this paper, we presented \emph{WetLinks:} a large-scale longitudinal Starlink dataset. It consists of six months of orchestrated Starlink measurements from two European cities: Osnabr\"{u}ck (DE), and Enschede (NL), totaling approximately 140,000 measurements.
It includes all relevant network parameters as well as accurate weather data captured with reference weather stations placed directly next to the Starlink dishes.
%Our measurement setup was special in the sense that the two measurement sites had a relatively close physical proximity, with a linear distance of approximately 80~km. This means that both were connected to the 53° orbit, and likely connected to the same satellites in sequence. Furthermore, we collected high-accuracy weather data using reference weather stations placed directly next to the Starlink dishes.
%At the time of publication, our dataset will consist of approximately 140,000 measurements. We made the dataset publicly available, making it, to the best of our knowledge, the largest publicly available Starlink dataset of its kind.

Based on our dataset, we analysed Starlink performance, including its susceptibility to weather conditions.
%By replicating the results of earlier smaller-scale studies, we validated our dataset.
We found that the download throughput varies throughout a day and drops in the afternoon. %This is likely due to load-spikes in the Starlink network, leading to reduced performance of standard traffic.
%Our analysis indicates no timely correlation over the course of a day to the upload throughput, RTT, and packet loss.
Moreover, we found that rain has a significant impact on the download throughput. First tests also indicate that clouds interfere with the signals. 
However, a more in-depth analysis with additional information about cloudiness is required to confirm this influence.
Generally, our results replicate earlier, smaller-scale studies. This shows the consistency of our dataset, enabling others to confidently build on it.

To enable the research community and practitioners to use our dataset in a plug-and-play fashion, we also release companion tooling that merges and preprocesses the different data sources (performance measurements, weather data, ...).
%Moreover, we published our analysis scripts to allow a full replicability of our results.

%We hope that our dataset helps networking researchers and practitioners to build AI models for their problems based on realistic network conditions from real-world measurements, while avoiding overfitting to certain network conditions.

\clearpage
% RvRD: no acknowledgements in the submission due to double-blind review
%\input{Acknowledgment}

\balance
\bibliographystyle{IEEEtranS}
\bibliography{IEEEabrv, bibliography}

\end{document}